\documentclass[a4paper,10pt]{article}
\usepackage{graphicx}
\usepackage{hyperref}

\usepackage{amsthm,amsmath,amssymb}
\usepackage[utf8]{inputenc}
\usepackage{authblk}
\usepackage{natbib}
\usepackage{tabularx}
\usepackage{booktabs}

  \newcolumntype{R}[1]{>{\raggedleft\let\newline\\\arraybackslash\hspace{0pt}}m{#1}}
\usepackage{threeparttable}

\usepackage{cleveref}
\usepackage[titletoc,title]{appendix}

\title{Is social capital associated with synchronization in human communication? An analysis of Italian call records and measures of civic engagement}

\author[d,*]{Marco Mamei}
\author[d]{Francesca Pancotto} 
\author[a,b]{Marco De Nadai}
\author[b,*]{Bruno Lepri}
\author[c]{Michele Vescovi}
\author[d]{Franco Zambonelli}
\author[e]{Alex Pentland}

\affil[a]{University of Trento, Trento, Italy}
\affil[b]{Fondazione Bruno Kessler, Trento, Italy}
\affil[c]{SKIL-TIM, Trento, Italy}
\affil[d]{University of Modena and Reggio Emilia, Modena, Italy}
\affil[e]{Massachusetts Institute of Technology, Cambridge, USA \newline {\small $^*$Corresponding authors:  marco.mamei@unimore.it, lepri@fbk.eu}}

\begin{document}

\maketitle

\begin{abstract} % abstract
Social capital has been studied in economics, sociology and political science as one of the key elements that promote the development of modern societies. It can be defined as the source of capital that facilitates cooperation through shared social norms. In this work, we investigate whether and to what extent synchronization aspects of mobile communication patterns are associated with social capital metrics. Interestingly, our results show that our synchronization-based approach well correlates with existing social capital metrics (i.e., Referendum turnout, Blood donations, and Association density), being also able to characterize the different role played by high synchronization within a close proximity-based community and high synchronization among different communities.
Hence, the proposed approach can provide timely, effective analysis at a limited cost over a large territory.
\end{abstract}

\section*{Introduction}
Synchronization is a process that allows the automatic coordination of units and events in time. Across many domains in nature, it is a mechanism that permits to reduce uncertainty and risk without the need for a centralized mechanism of control. Synchronization is a widespread phenomenon observed everywhere in nature, from animals~\cite{Sumpter2006} to neurons~\cite{Schneideman2006} and heart cells~\cite{Strogatz2003}, and up to more complex entities like human beings~\cite{Neda2000,Saavedra2010}. 

In humans, synchronization emerges as a spontaneous coordination mechanism that provides benefits to groups and the individuals that live within~\cite{HongPage2004}. 
In an evolutionary perspective, synchronization increases the probability of group survival, by reducing the individual costs required by the engagement of coordinated and cooperative action~\cite{Nowak2006}: in a multilevel selection mechanism, a group of cooperators has indeed higher chances of evolutionary success than a group of defectors. The positive effect of synchronization is also found in the behavior of people within groups, where synchronous activity has been found to enhance the level of cooperativeness~\cite{WiltermuthHeath2009} even without \textit{muscular bonding}~\cite{McNeill1997} or \textit{shared positive emotions}~\cite{Hannah1977,Ehrenreich2007}.
Synchronized groups should then in principle be more cooperative ones, and by comparing the level of synchronization between different groups, we may be able to measure their relative level of cooperativeness.
In the present study, we propose two \emph{synchronization indices}: (i) \emph{within synchronization} representing the relative level of cooperation within a close proximity-based community (i.e., municipality level), and (ii) \emph{between synchronization} representing the level of cooperation among different communities in a larger geographical area (i.e., province level).
More specifically, these indices capture the synchronization of human activity in an area through mobile phone data. Mobile phone data capture rich information about human activities and the structure of the social interactions therein~\cite{Schlapfer2014}.
They have been used to estimate the socioeconomic status of territories~\cite{Eagleetal2010} and individuals~\cite{Blumenstock2015}, to analyze the dynamics of cities~\cite{deNadai2016}, to model the spreading of diseases~\cite{wesolowski2012quantifying}, and to predict crime levels~\cite{Bogomolov2015}. Our hypothesis is that the two synchronization indices, capturing the degree of cooperativeness among human activities, can describe traditional measures of social capital, which is the source of capital that facilitates cooperation through shared social norms~\cite{Fukuyama2001}.

The relevance of social capital for economic growth is largely acknowledged~\cite{Putnam1993}; it reduces the transaction costs associated with formal coordination mechanisms,~\cite{KnackKeefer1997} predicts strong economic performance~\cite{Fukuyama1995} and financial development~\cite{Guisoetal2004}, and reduces corruption by inducing political and civic participation~\cite{Banfield1958,Nannicini2013}.

An important distinction in the social capital literature is the one between {\em bonding} and {\em bridging} patterns of relations~\cite{Putnam2000}. In his work, the political scientist Putnam states that {\em bonding social capital} provides emotional support and a sense of belonging in which the members of a community sustain each other \cite{Putnam2000}. This form of social capital is usually observed in homogeneous groups with strong cooperation, such as families or circles of close friends. {\em Bridging social capital}, instead, stems from relations between groups, that is, between individuals from heterogeneous backgrounds \cite{Putnam2000}. A community exploring novel interactions and co-operation with other communities can be considered to have a high amount of bridging social capital \cite{woolcock2000social}. This form of social capital has been described as potentially useful for achieving instrumental goals since a larger variety of resources becomes available by interacting with people of diverse status, occupation or ethnicity \cite{woolcock2000social}.

Previous research on capturing bonding and bridging social capital, and their effect on economic prosperity, from mobile phone and social media data has analyzed this issue focusing on the role played by different network structural properties (e.g., topological network diversity, network density, etc.)  \cite{Eagleetal2010,Norbutas2018}. To the best of our knowledge, the current work is the first study that analyzes whether and to what extent synchronization aspects of human communication are associated with traditional social capital metrics (i.e., Referendum turnout, Blood donations, and Association density).

Several studies have highlighted the role and the benefits played by the synchronization of activities among individuals and groups. 
Indeed, synchronization is argued to improve cooperation and trust in a community \cite{WiltermuthHeath2009,Saavedra2010}. Hence, we expect that communities with strong synchronization may experience richer opportunities for cooperation, decreased costs of market interactions, less reliance on formal business regulations and increased informal money circulation and investments, all aspects enabled by high levels of trust  \cite{Whiteley2000,WiltermuthHeath2009,Saavedra2010}. Thus, our {\em first hypothesis} is that high levels of call activity's synchronization in a tight area (that we associate to a municipality) are likely to reflect {\em bonding} patterns as people interact and communicate within a close proximity-based social group. In particular, high levels of {\em within synchronization} in a proximity-based community capture frequent communication patterns and connections among people living in this community.

Interaction among diverse groups of individuals and communities have been linked to higher exploration of possibilities, thus promoting the flow of information and novel ideas that affect economic prosperity~\cite{Schneideman2006,HongPage2004}. Following Paxton \cite{Paxton1999}, {\em bridging} social capital occurs when members of one group connect with members of other groups to seek access, support or to gain information. On this basis, our {\em second hypothesis} is that the interaction of a given community (i.e., a given municipality) with many different communities can be found in the high synchronization of their communication patterns. In particular, we expect that municipalities with more synchronization with other municipalities may experience a communication with a more diverse array of communities (i.e., having bridging ties spreading to many different municipalities) and gain novel ideas and information, and thus may show higher levels of {\em bridging} social capital.

Interestingly, our results show that a synchronization-based approach well correlates with traditional social capital measures (i.e., Referendum turnout, Blood donations, and Association density), being also able to characterize the different role played by high synchronization within a close proximity-based community and high synchronization among different communities.

\section*{Materials and methods}
For this study we use an aggregated and anonymized Call Detail Records (CDRs) dataset provided by the largest Italian mobile phone operator (34\% of market share) over a period of one month: from March 31, 2015 to April 30, 2015.
CDRs are collected for billing purposes by mobile network operators: every time a phone interacts with the network, a CDR recording the time and location (in terms of cell network's antenna) of the user is created\footnote{For a given phone call or SMS exchange we record only the CDR from the originating mobile terminal.}. The data we use is spatially aggregated and completely anonymized by the mobile phone operator as it is not possible to connect different calls of the same user.

Italy is an ideal playground in this domain because Italian regions present very different levels of economic development, although they have experienced the same formal institutions, laws, language and currency for many years now. Many scholars have identified the root of this persistent divergence in differential endowments of social capital~\cite{Bigonietal2016,Guisoetal2010}.
For these reasons, Italy has been widely studied in social capital economic literature~\cite{Banfield1958,Putnam2000}.
As a byproduct, there are several survey-based data sources for obtaining social capital measures that can be used as a ground-truth. More specifically, following examples in the economics literature \cite{Putnam2000,Guisoetal2004,Guisoetal2009}, we use \emph{Referendums turnout}, \emph{Association density} and \emph{Blood donations} as our ground-truth.
\emph{Referendums turnout} are usually considered as proxy of the desire of civic participation, as voting at referendums is not mandatory in Italy and the issues on the ballot in referendums are less related to local interests.
\emph{Association density} is defined as the number of associations per 100,000 inhabitants. Associations can be cultural, leisure, artistic, sports, environmental, and any kind of nonprofit associations with the exclusion of professional and religious associations~\cite{Putnam1993}.
\emph{Blood donations} are measured as the instances of donations per 1,000 inhabitants.

In our analysis, we select both large provinces (NUTS-3 regions) with more than one million inhabitants, and smaller provinces known for high and low levels of social capital (according to the aforementioned social capital survey-based measures). The indicators of level of social capital used to select small NUTS-3 regions - intended with a population between 200,000 and 500,000 inhabitants - are the data available for Italy on association density, referendum participation and blood donations~\cite{Bigonietal2016,Buonannoetal2009, Cartocci2007}. Specifically, considered NUTS-3 regions are:

\begin{itemize}
\item Turin, Milan, Venice, Rome, Naples, Bari, Palermo (large NUTS-3 regions);
\item Caltanissetta, Siracusa, Benevento, Campobasso (defined as low-social capital NUTS-3 regions~\cite{Cartocci2007});
\item Siena, Ravenna, Ferrara, Asti, Modena (defined as high-social capital NUTS-3 regions~\cite{Cartocci2007}).
\end{itemize}

\begin{figure}[ht!]
    \centering
    \includegraphics[width=0.9\columnwidth]{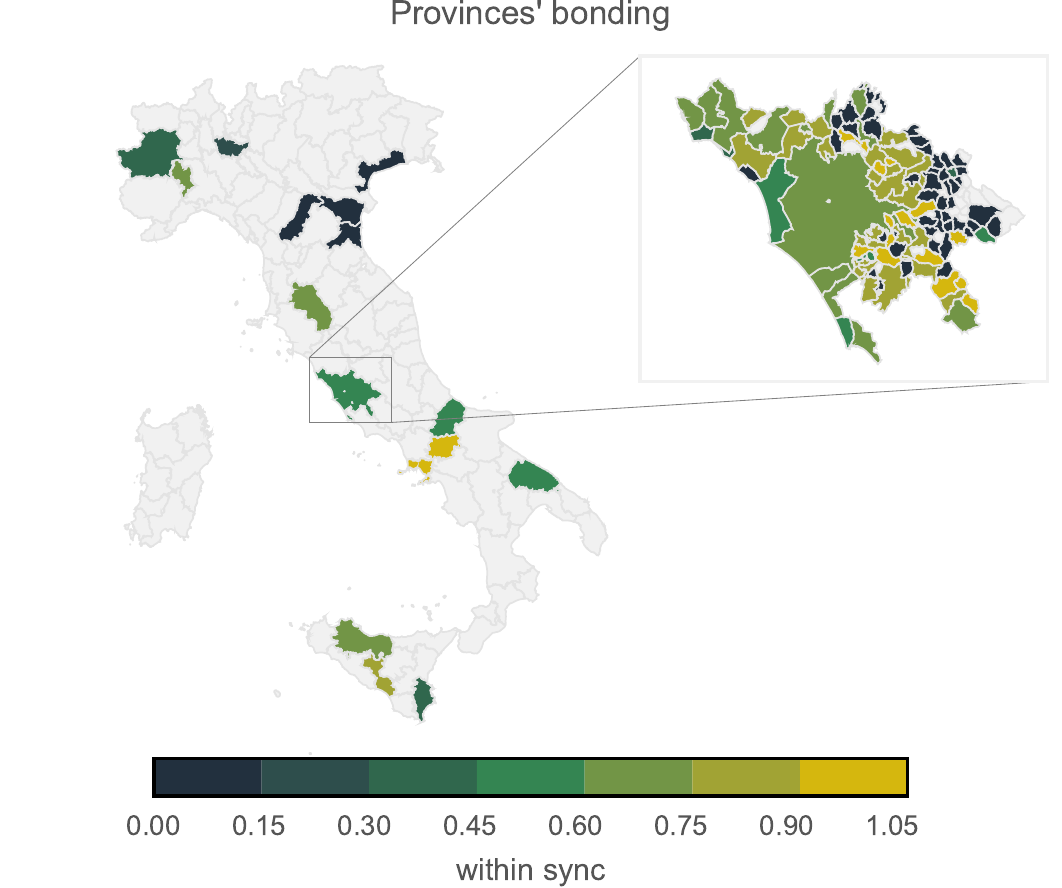}
    \label{fig:map}
    \caption{Analyzed data from large NUTS-3 regions ($>1M$ inhabitants), and medium NUTS-3 regions known for high/low levels of social capital~\cite{Cartocci2007}. {\bf (right inset)} Enlargement of Rome NUTS-3 region highlighting municipalities (LAU-2 regions). Data are collected at a sub-municipality resolution.}
\end{figure}

These areas represent the smallest areal units available for social capital data. NUTS-3 regions are therefore our unit of analysis. The choice of these NUTS-3 regions is partly data-driven, but we select them also as they exhibit different levels of social capital. Figure \ref{fig:map} shows the map of Italy with the NUTS-3 regions under analysis.

The area of each region is spatially divided in an irregular grid, provided by the mobile phone operator, based on the size of the underlying antennas' coverage area.
The cells have area ranging from 0.04 $Km^2$ in the city center to 40 $Km^2$ in the suburbs.

For each cell, we aggregate the number of CDRs at an hourly time scale to obtain a time series recording the level of activity on an hourly basis.

We normalize each $i$-th cell's time series $x^i_{t=day,h}$ with a z-score computed on an hourly basis. $\mu^i_h$ and $\sigma^i_h$ are the 24 means and standard deviations of $x^i_{day,h}$ for each hour.
Thus, we obtain: $z^i_{day,h} = (x^i_{day,h} - \mu^i_h) / \sigma^i_h$.
Using different $\mu^i_h$ and $\sigma^i_h$ for different hours is very important because otherwise the circadian trend in our data would notably bias the synchronization among the time series (i.e., all time series would be highly synchronized because the day-night trend would cover more subtle differences).

\begin{figure}
    \centering
    \includegraphics[width=0.9\columnwidth]{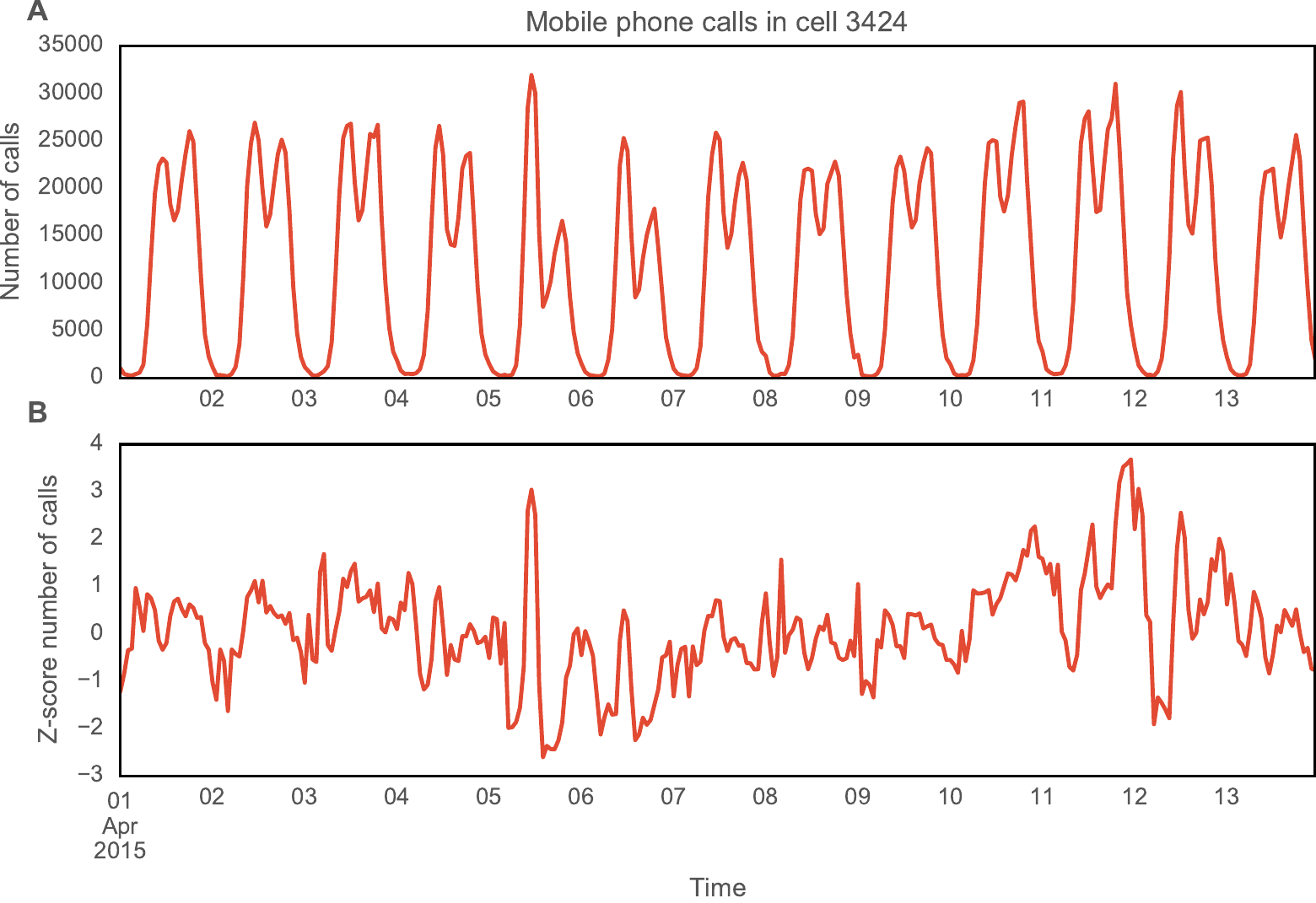}
    \label{fig:dataplot}
    \caption{Example of daily rhythm in a mobile phone cell. A) Original behaviour extracted from mobile phone data. B) Z-score scaled behaviour extracted from mobile phone data.}
\end{figure}

The resulting time series (see Figure \ref{fig:dataplot}) highlights deviations of the mean activity in different hours of the day on the one hand and on the other they are sufficiently stationary to apply standard statistics to measure the correlation (i.e., synchronization) of two time series.

For each NUTS-3 region, we compute two synchronization metrics: {\em within synchronization} is the average daily synchronization among cells assigned to the same municipality; {\em between synchronization} is the average daily synchronization among cells assigned to different municipalities (cells are assigned to municipalities based on the quantity of their overlapping area).
Specifically, for each couple of cells $i$ and $j$, we compute the average daily Mutual Information between $z^i_{day,h}$ and $z^j_{day,h}$: $\frac{1}{N}\sum_{day=1}^N I(z^i_{day,h};z^j_{day,h})$.

Mutual information is a natural measure of non-linear dependence quantifying the amount of information obtained about one time-series through the other one.
Therefore, it measures how synchronized the two series are, and it is computed as:
$$
I(z^i_{day,h};z^j_{day,h}) = \int_{z^i_{day,h}} \int_{z^j_{day,h}}  p(z^i_{day,h},z^j_{day,h})log\left(\frac{p(z^i_{day,h},z^j_{day,h})}{p(z^i_{day,h})p(z^j_{day,h})}\right)
$$

This approach computes a single average ({\em within} and {\em between}) synchronization for the whole time of observation (one month with our data). So, even if short-term events can spur sudden synchronization, the average value reflects longer-term trends in the behavioral patterns in the regions.

\begin{figure*}[ht!]
    \centering
    \includegraphics[width=0.9\textwidth]{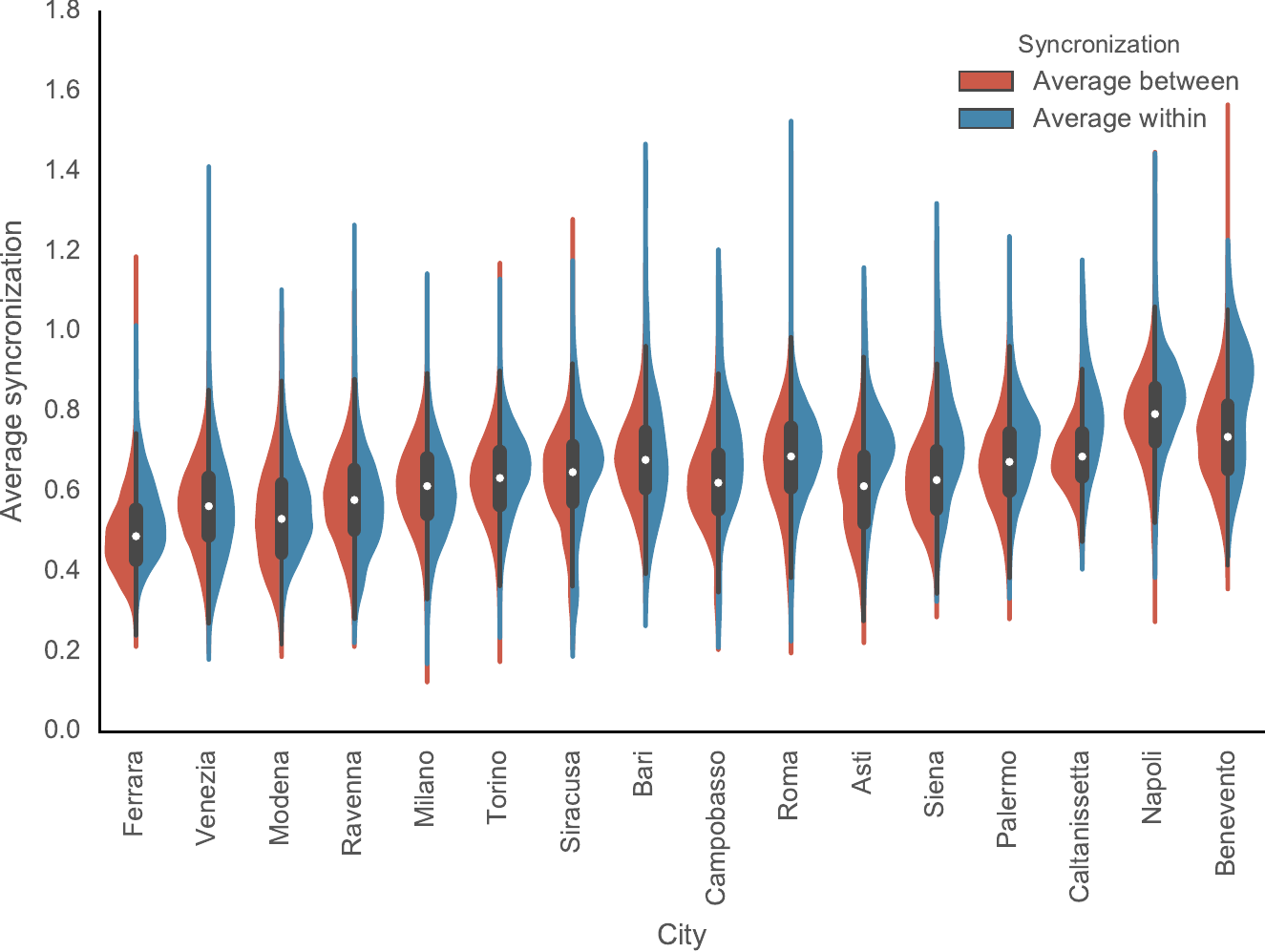}
    \label{fig:violins}
    \caption{Violin plots, ordered by the median \emph{within synchronization}, showing the average \emph{between} and \emph{within synchronization} of each city.}
\end{figure*}

Figure \ref{fig:violins} shows the distribution of \emph{between} and \emph{within} synchronization for the NUTS-3 regions under analysis.
We consider the mean (among cells) of \emph{between} and \emph{within} synchronization as the reference value for each region (to be used in the regression model described below).

As aforementioned in the Introduction Section, we postulate that:
\begin{itemize}
\item High levels of {\em within synchronization} reflect the tendency of people to communicate together within their spatial cluster (i.e., municipality).
\item High levels of {\em between synchronization} reflect instead the tendency of people to communicate together across different spatial clusters (i.e., municipalities).
\end{itemize}

We therefore use these two synchronization measures, computed from passively collected human behavioural data, to describe traditional proxies for social capital used in economics literature such as \emph{Referendums turnout}, \emph{Association density} and \emph{Blood donations}.

In summary, for each of the 16 NUTS-3 regions under analysis, we compute the respective synchronization indices (i.e., {\em within} and {\em between} synchronization) and extract the traditional proxies for social capital. We check via Moran's I test that the obtained variables are not spatially auto-correlated, then we apply the linear regression analysis described in the following section.

\subsection*{Regression analysis.}

To validate our hypotheses, we describe the three social capital measures (i.e., Referendums turnout, Blood donations, and Association density) by means of three Ordinary Least Squares (OLS) models where the independent variables are: (i) \emph{within synchronization}, (ii) \emph{between synchronization}, and (iii) per-capita income.
In principle many factors could affect the level of social capital and thus affect our estimation: the quality of institutions, the level of education, the degree of income inequality, to mention some.
Following Alesina \emph{et al.}~\cite{AlesinaLaferrara2002} and Guiso \emph{et al.}~\cite{Guisoetal2016} we here consider per-capita income as a sole co-variate for the regression, to keep our estimates parsimonious, and use the level of per-capita income as a general proxy for these factors. Indeed higher per-capita income has been shown to be related to the strength of local institutions~\cite{HelliwellPutnam1995} and to the quality of education systems~\cite{Fukuyama2001}.
In Appendix C we report an additional set of regression analyses using the fraction of illiterate population, a good proxy for the level of education, as a sole covariate for the regression.

\emph{Between} and \emph{within synchronization} across NUTS-3 regions are highly correlated ($\rho = 0.9$), raising multicollinearity issues.
Having correlated regressors, we have to rely on multiple metrics to illustrate the statistical significance and importance of the variables in our model~\cite{nathans2012interpreting}.
Thus, we report and discuss the variable importance through the beta weights, structure coefficients~\cite{courville2001use}, commonality analysis components~\cite{rowell1996}, dominance analysis~\cite{azen2003dominance} and Lindeman, Merenda, and Gold's (LMG) method~\cite{lindeman1980introduction}.

Beta weights are often relied on to assess regressors' importance~\cite{courville2001use}. Beta weights indicate the expected increase/decrease  in the dependent variable (e.g., Referendums turnout), expressed in standard deviation units, given a one standard deviation increase in such independent variable with all other independent variables held constant. However, the sole reliance on beta weights to interpret the contribution of each independent variable is justified only when the independent variables are perfectly uncorrelated~\cite{pedhazur1997}.
In fact, beta weights may receive credit for explained variance shared with other regressors, while beta weights of the other regressors are not given credit for this shared variance~\cite{pedhazur1997}. Therefore, the contribution of the other regressors to the regression effect may be not fully captured.
Moreover, beta weights have also limitations in determining \emph{suppression} effects in a regression, that is, a regressor that contributes little or no variance to the dependent variable but it may have a large non-zero beta weight because it \emph{purifies} one or more regressors of their irrelevant variance, thereby increasing its or theirs predictive power~\cite{capraro2001}.

Structure coefficients quantify the strength of the bi-variate relationship between each regressor and the dependent variable in isolation from other correlations between regressors and dependent variable. Hence, they are a useful measure of the \emph{direct} effect of a regressor~\cite{courville2001use}.
Being only a measure of direct effect, they are unable to identify regressors sharing explained variance in the dependent variable, and thus to quantify the amount of this shared variance~\cite{courville2001use}. Instead, the LMG measure can be thought as the average improvement of regressor $X_1$, over all models of size $s$ without $X_1$~\cite{lindeman1980introduction}.

In order to quantify  the  contribution that each regressor shares with every other set of regressors, we also perform a commonality analysis~\cite{rowell1996}. This technique decomposes $R^2$, and thus the total effect ($Tot_{CA}$), into its \emph{unique} ($U_{CA}$) and \emph{common} ($C_{CA}$) effects.
Unique effects indicate how much variance is uniquely accounted for by a single regressor; while common effects indicate how much variance is common to each set of regressors~\cite{rowell1996}. It is worth noting that if the regressors are all uncorrelated, the contributions of all regressors are unique effects, as no variance is shared between independent variables in the prediction of the dependent variable.

Moreover, we use dominance analysis~\cite{azen2003dominance} to determine the importance of a regressor based on comparisons of unique variance contributions of all pair of independent variables to regression equations involving all possible subsets of regressors. Interestingly, dominance analysis is a technique able to quantify (i) the \emph{direct} effect of a regressor in isolation from other regressors, as the subset containing no other regressors includes zero-squared correlations, (ii) the \emph{total} effect, as it compares the unique variance contributions of the regressors when all of them are included in the model, and (iii) the \emph{partial} effect, as it compares the unique variance contributions of the regressors for all the possible subsets of them.

\section*{Results}

Results of OLS models are shown in Table~\ref{tab:regressions}, where we report the adjusted $R^2_{adj}$\footnote{The adjusted $R^2_{adj}$ is a variant of the $R^2$ that aims at overcoming the spurious increase of the former when extra variables are added to the model. It is defined as $R^2_{adj}=1-(1-R^2)\frac{n-1}{n-k-1}$ where $n$ is the number of data-points and $k$ the number of parameters in the model.} of the OLS using \emph{between} synchronization, \emph{within} synchronization and per-capita income as covariates.

\begin{figure*}[ht!]
    \centering
    
    \includegraphics[width=0.9\textwidth]{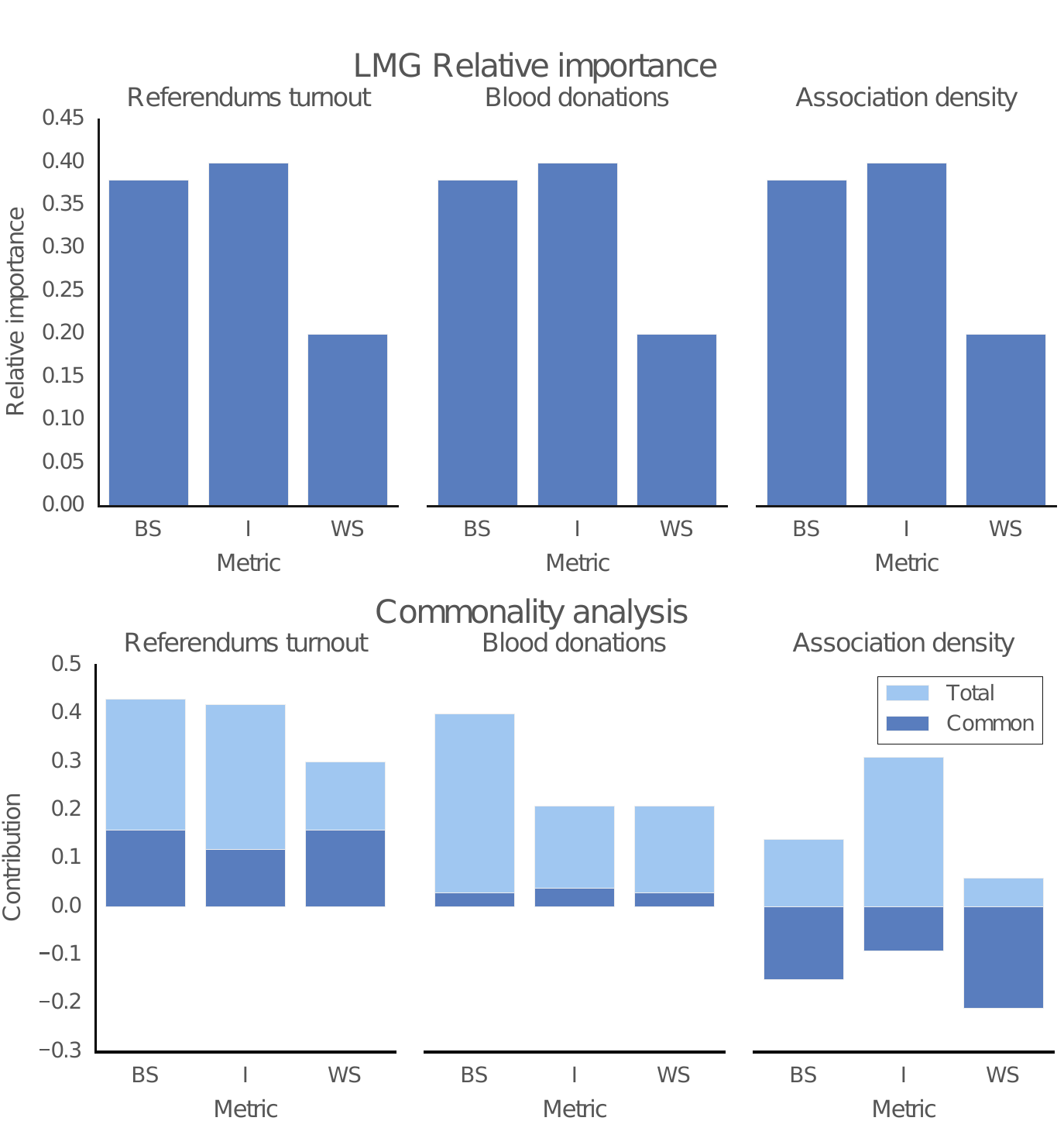}
    \caption{upper) Lindeman, Merenda and Gold relative importance of the independent variables we used in our model; lower) Total, common and unique contribution of the independent variables we used in our model. (BS): \emph{between synchronization}. (I): per-capita income. (WS): \emph{within synchronization}.}
    
    \label{fig:importance}
\end{figure*}

The variable importance of the independent variables is reported through the Beta weights, the structure coefficients~\cite{courville2001use}, the commonality analysis components~\cite{rowell1996}, the dominance analysis~\cite{azen2003dominance} and the Lindeman, Merenda, and Gold's (LMG) method~\cite{lindeman1980introduction}. \Cref{fig:importance} summarizes the results of two of the most used variable importance metrics.

Here we provide a detailed analysis of each social capital proxy used in economics literature.

\begin{table}[ht]
   \centering
    \footnotesize

    \begin{tabularx}{\columnwidth}{@{}Xrrrrrr@{}}
        \toprule
        & $\mathbf{\beta}$ \textbf{(}$\mathbf{95\%}$ \textbf{CI)} & $\mathbf{r_s}$ & $\mathbf{U_{CA}}$ & $\mathbf{C_{CA}}$ & $\mathbf{Tot_{CA}}$ & LMG
        \\
        \midrule
        \textbf{\textit{Referendums turnout}} \\($\mathbf{R^2_{adj}: 0.68}$)\\
        between sync\tnote{s} 	& -0.12** (-0.20, -0.05) & -0.76 & 0.27 & 0.16 & 0.43 & 0.38  \\
        within sync\tnote{s}  & 0.09* (0.01, 0.18) & -0.63 & 0.13 & 0.16 & 0.30 & 0.20 \\
        per-capita income & 0.06** (0.02, 0.10) & 0.75 & 0.30 & 0.12 & 0.42 & 0.40 \\

        \midrule
        \textbf{\textit{Blood donations}}  \\($\mathbf{R^2_{adj}: 0.55}$)\\
        between sync\tnote{s} 	& -24.91** (-40.44,    -9.37) & -0.79 & 0.36 & 0.03  & 0.40 & 0.52  \\
        within sync\tnote{s}  & 19.49* (2.45, 36.54) & -0.58 & 0.18 & 0.03 & 0.21 & 0.24 \\
        per-capita income & 8.49* (0.67, 16.31) & 0.57 & 0.16 & 0.04 & 0.21 & 0.22 \\

        \midrule
        \textbf{\textit{Association density}} \\($\mathbf{R^2_{adj}: 0.52}$)\\
        between sync\tnote{s} 	& -21.88** (-37.54, -6.23) & -0.48	& 0.29   & -0.15  & 0.14 & 0.30  \\
        within sync\tnote{s}  & 22.96* (5.78, 40.14) & -0.31 & 0.27 & -0.21 & 0.06 & 0.27 \\
        per-capita income & 13.00** (5.12, 20.88) & 0.71 & 0.41 & -0.09 & 0.31 & 0.42 \\
         \bottomrule
    \end{tabularx}
     \caption{Referendums turnout, Blood donations, Association density represented by \emph{between} and \emph{within synchronization}, controlled for per-capita income were tested using commonality analysis. As for statistical significance of the beta weights, we use the following notation: *$p<0.05$, **$p<0.01$.}
    \label{tab:regressions}
\end{table}

\mbox{ } \\
\textbf{Referendums turnouts.} The first group of rows of Table~\ref{tab:regressions} shows that \emph{between synchronization} contributes the most to the regression equation ($\beta = -0.12$), while holding all other regressors constant.
It is the most correlated variable with the predicted \emph{Referendums turnout} ($r_s = -0.76$) and the major contributor to the regression effect ($Tot_{CA} = 0.43$), where $27.2\%$ of regression effects is \emph{unique} and $16.2\%$ is in common with the other variables.
The relative importance of \emph{between synchronization} ($Tot_{CA} = 0.43$ and $LMG = 0.38$) is closely related to the one of \emph{per-capita income} ($Tot_{CA} = 0.42$ and $LMG = 0.40$).

The second most important beta weight is \emph{within synchronization} that, besides its positive value, has negative correlation with \emph{Referendums turnout} ($r_s = -0.63$).
This may indicate that the regression effect was confounded by all the variables included in the model but they all contribute substantially in the explanation of \emph{Referendums turnout} (all $C_{CA}$ and $Tot_{CA}$ values are greater than zero).

\begin{table}[ht]
    \centering
    \small
    \begin{tabularx}{\columnwidth}{@{}Xccc@{}}
        \toprule
         \textbf{Dominance} & \textbf{Complete}& \textbf{Conditional} & \textbf{General} \\ \midrule
        \emph{between sync} $>$ \emph{within sync} & $\checkmark$ & $\checkmark$ & $\checkmark$ \\
        \emph{between sync} $>$ \emph{per-capita income} &&&$\times$\\
         \emph{within sync} $>$ \emph{per-capita income}  & $\times$ & $\times$ & $\times$ \\
         \bottomrule
    \end{tabularx}
    \caption{Referendums turnout: Dominance analysis output. The $\checkmark$ symbol represents the dominance of a variable $A$ on $B$. The $\times$ symbol represents the dominance of a variable $B$ on $A$. In empty cells dominance could not be established between regressors. }
    \label{tab:dominance_referenda}
\end{table}

\mbox{ } \\
\textbf{Blood donations.} From the second group of rows of Table~\ref{tab:regressions} we observe that \emph{between synchronization} holds the highest contribution to the regression in all the metrics, accounting for $52\%$ of the importance in the model, ($\beta = -24.91$), highest \emph{total} ($Tot_{CA} = 0.40$) and \emph{unique} contribution ($U_{CA} = 0.36$).

The second most important beta weight is \emph{within synchronization} that, besides its positive value, has negative correlation with \emph{Blood donations} ($r_s = -0.580$).
This may indicate that the regression effect was confounded by all the variables included in the model but they all contribute substantially in the explanation of \emph{Blood donations} (all $C_{CA}$ and $Tot_{CA}$ values are greater than zero).
The importance of \emph{within synchronization} is very close to the importance of \emph{per-capita income}, but from the Dominance analysis (see \Cref{tab:dominance_blood}) we have that \emph{per-capita income} has a minor role in the regression.

\begin{table}[ht]
    \centering
    \small
    \begin{tabularx}{\columnwidth}{@{}Xccc@{}}
        \toprule
         \textbf{Dominance} & \textbf{Complete}& \textbf{Conditional} & \textbf{General} \\ \midrule
        \emph{between sync} $>$ \emph{within sync} & $\checkmark$ & $\checkmark$ & $\checkmark$ \\
        \emph{between sync} $>$ \emph{per-capita income} & $\checkmark$ & $\checkmark$ & $\checkmark$\\
         \emph{within sync} $>$ \emph{per-capita income}  & $\checkmark$ & $\checkmark$ & $\checkmark$ \\
         \bottomrule
    \end{tabularx}
    \caption{Blood donations: Dominance analysis output. The $\checkmark$ symbol represents the dominance of a variable $A$ on $B$. }
    \label{tab:dominance_blood}
\end{table}

\mbox{ } \\
\textbf{Associations density.} The last group of rows in Table~\ref{tab:regressions} shows that \emph{within synchronization} and \emph{between synchronization} obtained the largest beta weights ($\beta = 22.96$ and $\beta = -21.88$ respectively), demonstrating the most important contributions to the regression equation, while holding all other regressors constant.
Despite this, \emph{per-capita income} accounts for $42\%$ of the importance in the model, having also the highest \emph{total} ($Tot_{CA} = 0.42$) and \emph{unique} contribution ($U_{CA} = 0.41$).
From the Dominance analysis (see \Cref{tab:dominance_assoc}) it is possible to see that the most important variable is indeed \emph{per-capita income}, followed by \emph{between synchronization} and \emph{within synchronization}.

Particularly, besides the positive value of \emph{within synchronization}'s beta weight, it is negatively correlated with \emph{Association density} ($r_s = -0.31$).
Together, the very small structure coefficient ($r^2_s = 0.09$) and the negative common effect ($C_{CA} = -0.21$) may indicate~\cite{kerlinger1973multiple} the suppression role of \emph{within synchronization} in the regression that \emph{purifies} the variance explained by the other variables.

\begin{table}[ht]
    \centering
    \small
    \begin{tabularx}{\columnwidth}{@{}Xccc@{}}
        \toprule
         \textbf{Dominance} & \textbf{Complete}& \textbf{Conditional} & \textbf{General} \\ \midrule
        \emph{between sync} $>$ \emph{within sync} & $\checkmark$ & $\checkmark$ & $\checkmark$ \\
        \emph{between sync} $>$ \emph{per-capita income} & $\times$ & $\times$ & $\times$\\
         \emph{within sync} $>$ \emph{per-capita income}  & $\times$ & $\times$ & $\times$ \\
         \bottomrule
    \end{tabularx}
    
    \caption{Association density: Dominance analysis output. The $\checkmark$ symbol represents the dominance of a variable $A$ on $B$. The $\times$ symbol represents the dominance of a variable $B$ on $A$.}
    \label{tab:dominance_assoc}
\end{table}

\section*{Discussion}

\begin{figure*}[ht!]
    \centering
    
    \includegraphics[width=0.9\textwidth]{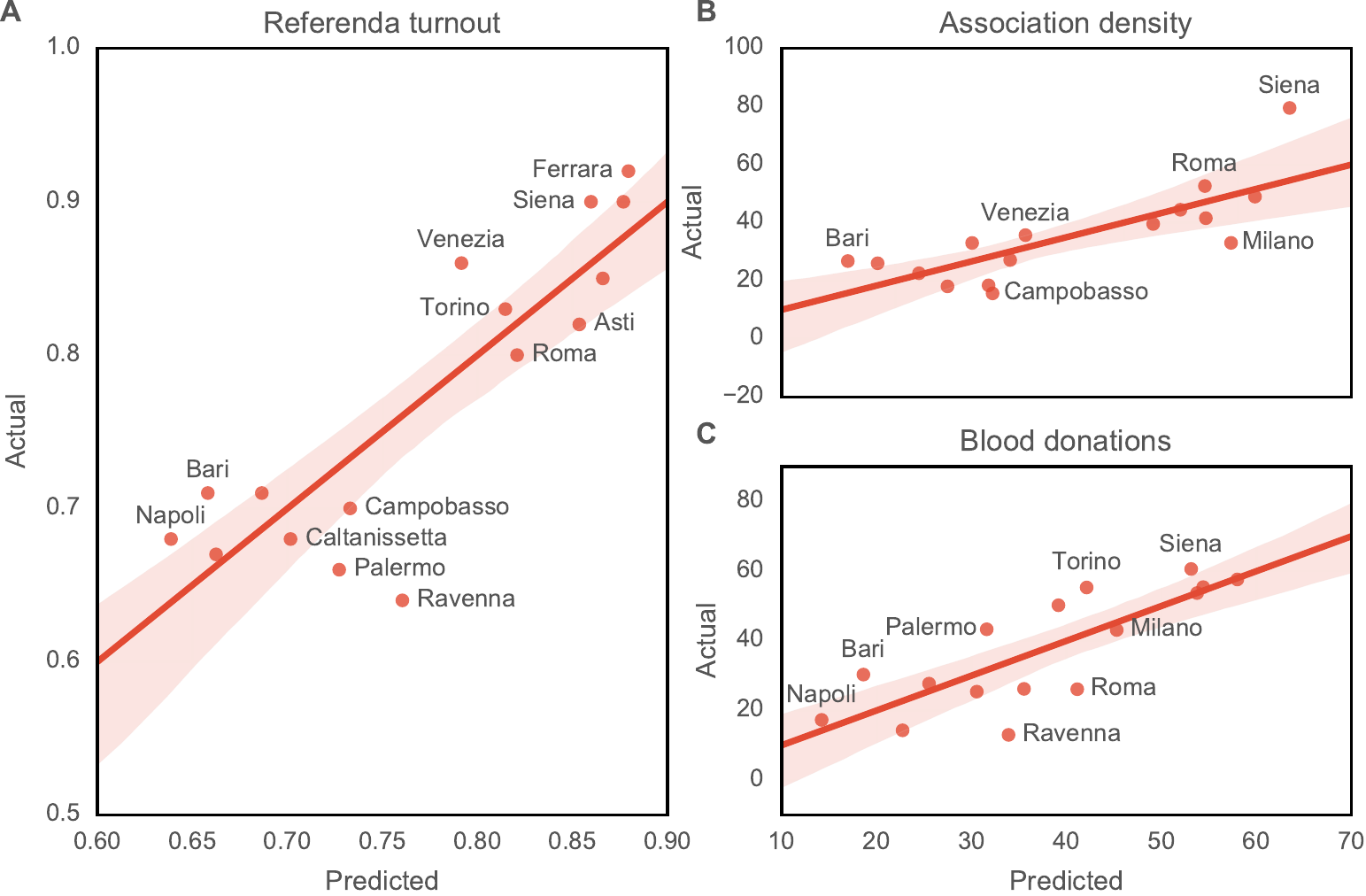}
    \caption{A) Relation between actual referendums turnout (as reported in the official ISTAT statistics) and predicted referendums turnout (as inferred from mobile phone data); B) Relation between actual association density and predicted association density; C) Relation between actual blood donations and predicted blood donations.}
    
    \label{fig:regressions}
\end{figure*}

Taken together, our results show that the models can explain the 68\% of the variation in \emph{Referendums turnout} ($R^2_{adj}$ = 0.68), the 55\% of the variation in \emph{Blood donations} ($R^2_{adj}$ = 0.55) and the 52\% of the variation in \emph{Association density} ($R^2_{adj}$ = 0.52).
\Cref{fig:regressions} shows the distribution of the fitted points.

Particularly, {\em within synchronization} correlates positively with social capital metrics ($\beta=0.09$ for Referendums turnout, $\beta=19.49$ for Blood donations, and $\beta=22.96$ for Association density).
Thus, this indicator informs us on the intensity of cohesion within close-proximity groups and communities, which approximates {\em ``... the instantiated informal norm that promotes co-operation between two or more individuals...~\cite{Fukuyama2001}".} 

In Larssen et al, individuals with strong social bonding (i.e., association and trust among neighbors) are more likely to take civic action.

Our second indicator, {\em between synchronization}, captures the tendency of a given community (i.e.,  a  given  municipality) to communicate with many  different  communities (i.e., other municipalities). Thus, more {\em between synchronization} implies more interaction among multiple groups (i.e., municipalities); while less {\em between synchronization} implies less interaction and more isolation among groups.
Interestingly, our results correlate negatively a high level of {\em between synchronization} with standard social capital metrics ($\beta=-0.12$ for Referendums turnout, $\beta=-24.91$ for Blood donations, and $\beta=-21.88$ for Association density). These findings are in line with a number of theoretical and empirical works claiming that diversity undermines a sense of community and social cohesion~\cite{KnackKeefer1997,AlesinaLaferrara2000,Glaeser2000,AlesinaLaferrara2002,costaKahn2002,MiguelGugerty2005}. For example, Alesina and La Ferrara~\cite{AlesinaLaferrara2000} have studied whether and how much the degree of heterogeneity in communities influences the amount of participation in different types of groups. Using survey data on group membership and data on localities in United States, they found that, after controlling for many individual characteristics, participation in associations (e.g., religious groups, hobby clubs, youth and sport groups, etc.) is significantly lower in more different, unequal, and racially or ethnically fragmented localities.

Our results are obtained including per-capita income in the regressions, similarly to what is done in the literature~\cite{AlesinaLaferrara2002,Guisoetal2004}; controlling for wealth at the level of the NUTS-3 regions.
The role of per-capita income is indeed important. We find that per-capita income has a strong relevance in describing the Association density, while it shows a minor role in explaining the higher Referendums turnout and Blood donations.

\section*{Conclusion}

In this paper, we have introduced a couple of novel synchronization metrics (i.e., {\em within} and {\em between synchronization}) that represent an innovative and efficient way to describe traditional social capital measures (i.e., Referendum turnouts, Blood donations, and Association density).
The proposed approach is, at the best of our knowledge, the first one that combines synchronization metrics and mobile phone data, which are always up to date and available for a very large fraction of the world population. 

A further merit of our approach is the ability to identify and analyze individually the role played by the level of
cooperation within a close proximity-based community (i.e., {\em within synchronization}), and the one played by the level of cooperation among different communities in a larger geographical area (i.e., {\em between synchronization}). Moreover, our approach does not need individual-level data, which is rarely shared by telecommunication operators to ensure data confidentiality. It is also worth noting that our synchronization-based approach can be extended easily to other sources of information such as activities on social media platforms, mobility routines captured from transportation data, etc.

Social capital is a key determinant to understand neighborhood stability for crime prevention, to enforce social cohesion, e.g., immigrant integration, and to create integration tools ind addition to language and culture training.
Thus, the geographical characterization of areas with differential levels of social capital is an important tool in the hands of policy makers aiming at specific incentive policies, which are clearly more or less effective depending on the underlying social capital types and levels.

\begin{appendices}

\section{Regression using Multiple Deprivation Index}

\begin{figure*}[ht!]
\centering
\includegraphics[width=0.8\textwidth]{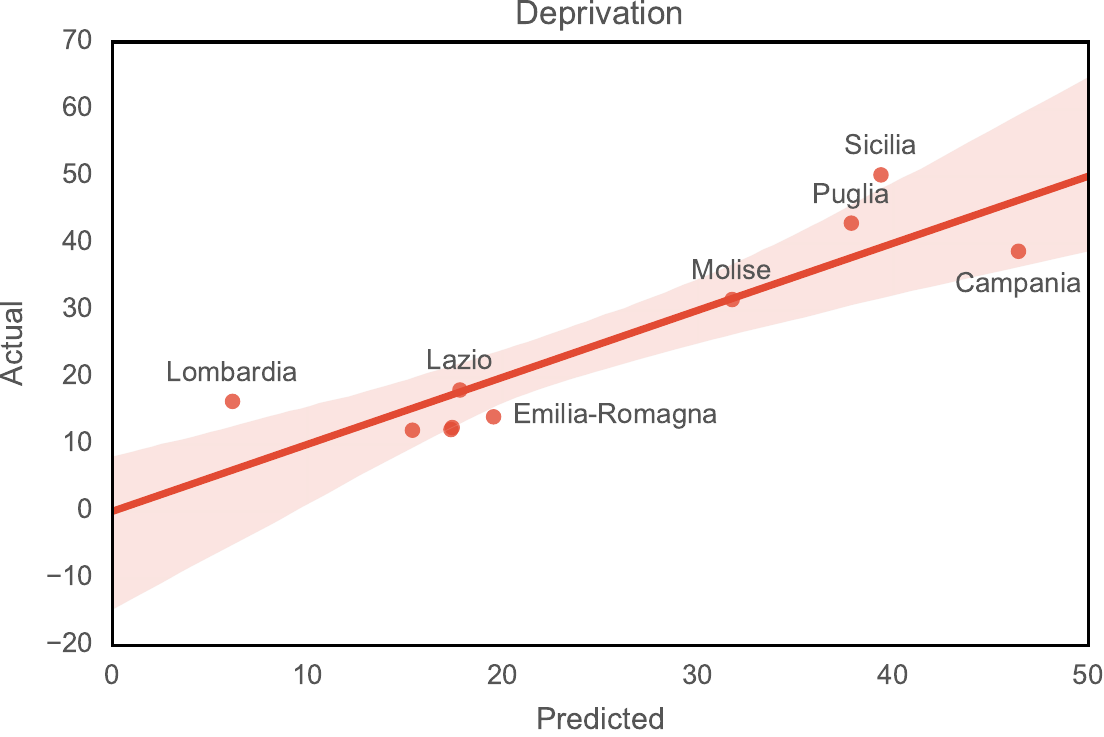}
\caption{Relation between actual deprivation index (as reported in the official ISTAT statistics) and predicted deprivation index (as inferred from mobile phone data).}
\label{fig:deprivation_regression}
\end{figure*}

While there seems to be a growing empirical evidence that social capital contributes significantly to sustainable development, a number of authors  raise issues and point to unconvincing and conflicting results \cite{Bigonietal2016,Sabatini2008}. At the heart of the problem is the multiple definitions and metrics of both social capital and sustainable development. Following this line of research and other similar works \cite{Eagleetal2010,Blumenstock2015}, we analyze the association between our synchronization metrics (i.e., {\em within} and {\em between synchronization}) and the Multiple Deprivation Index, see Figure \ref{fig:deprivation_regression}. Multiple Deprivation Index is a synthetic measure used for analyzing social exclusion. It combines information comprising household structure, level of education and participation in the labour market. Our data is based on official ISTAT statistics and refer to year 2013.

Having the deprivation data available only at the NUTS-2 region level, the regression is applied only to few data-points.
This issue causes high instability of the coefficients of the OLS regression (see $95\% CI$ column of Table~\ref{tab:analysis_depriv}). For this reason we show here the results of the analysis (Table~\ref{tab:analysis_depriv}) and the dominance results (Table~\ref{tab:dominance_assoc}) without deep explanations.
Nevertheless, the explained variance is very high, meaning that this associative relation should be further investigated in future studies.

\begin{table}[ht!]
    \centering
    \footnotesize
    
    \begin{tabularx}{\columnwidth}{@{}Xrrrrrrl@{}}
        \toprule
        
        & $\mathbf{\beta}$ ($\mathbf{95\%}$ CI) & $\mathbf{r_s}$ & $\mathbf{r^2_s}$ & $\mathbf{U_{CA}}$ & $\mathbf{C_{CA}}$ & $\mathbf{Tot_{CA}}$ & LMG \\ \midrule
        $\mathbf{R^2_{adj}: 0.68}$\\
        between sync\tnote{s} 	& 10.84 (-4.36,   26.05) & 0.76 & 0.58	& 0.10   & 0.35  & 0.46 & 0.27  \\
        within sync\tnote{s}  & -4.92 (-21.68,   11.83) & 0.76 & 0.58 & 0.01 & 0.44 & 0.46 & 0.19 \\
        per-capita income & -9.91 (-18.10, -1.75) & -0.85 & 0.73 & 0.30 & 0.27 & 0.57 & 0.52 \\
         \bottomrule
    \end{tabularx}
    
    \caption{Deprivation represented by \emph{between} and \emph{within synchronization}, controlled for per-capita income was tested using commonality analysis.}
    \label{tab:analysis_depriv}
\end{table}

\begin{table}[ht!]
    \centering
    \small
    \begin{tabularx}{\columnwidth}{@{}Xccc@{}}
        \toprule
         \textbf{Dominance} & \textbf{Complete}& \textbf{Conditional} & \textbf{General} \\ \midrule
         \emph{between sync} $>$ \emph{within sync} & $\checkmark$ & $\checkmark$ & $\checkmark$ \\
         \emph{between sync} $>$ \emph{per-capita income} & $\times$ & $\times$ & $\times$\\
         \emph{within sync} $>$ \emph{per-capita income}  & $\times$ & $\times$ & $\times$ \\
         \bottomrule
    \end{tabularx}
    \caption{Deprivation Dominance analysis output. The $\checkmark$ symbol represents the dominance of a variable $A$ on $B$. The $\times$ symbol represents the dominance of a variable $B$ on $A$.}
    \label{tab:dominance_depriv}
\end{table}

\section{Correlation Matrix Among Variables}

\begin{figure*}[ht!]
\centering
\includegraphics[width=0.8\textwidth]{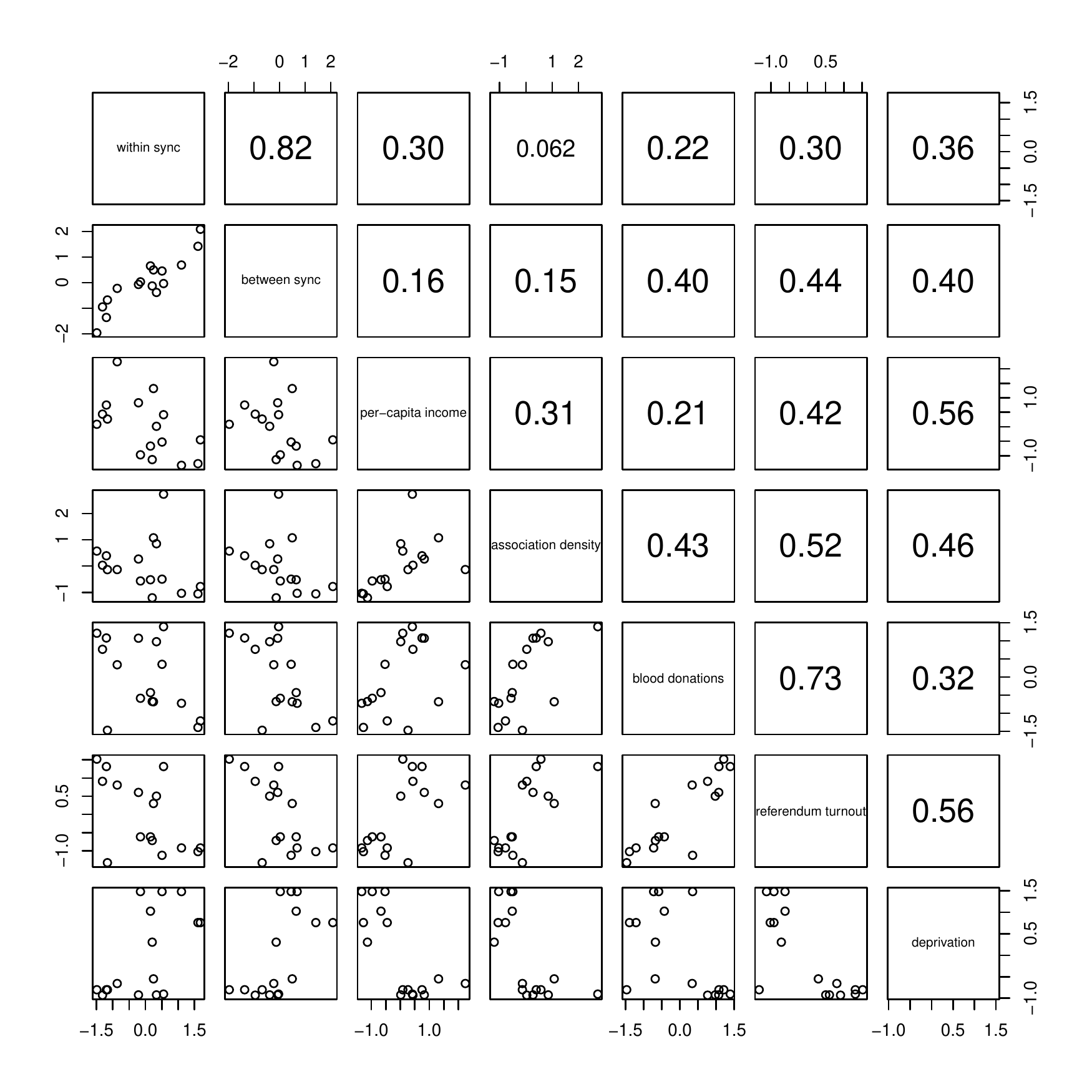}
\caption{Correlation matrix between all the variables. The upper panel reports $R^2$ among pairs.}
\label{fig:pairs}
\end{figure*}

To present the described correlation and dominance analysis in a more intuitive way, in Fig. \ref{fig:pairs}, we report the correlation matrix among all the variables. It is possible to see that $R^2$ among pairs is lower than in the multiple regression case. The per-capita income has an important role as confounding factor (and has been included into the covariates for this reason), but by no means it is able to explain the regression alone.

\section{Testing the Robustness}
We conduct some additional analyses to test the robustness of our approach.
Firstly, we verify the impact of the temporal aggregation used to compute the (\emph{within} and  \emph{between}) synchronization values. While in the main text, we use CDR counts aggregated using a 1-hour temporal window, in Table~\ref{table:2-hours} we also report the results obtained with a 2-hours temporal window. Results remain similar, but due to the limited number of data points we often lose statistical significance.

\begin{table}[ht]
    \centering
    \footnotesize

    \begin{tabularx}{\columnwidth}{@{}Xrrrr@{}}
        \toprule
        & \textbf{Referendum turnout} & \textbf{Blood donations} & \textbf{Association Density} 
        \\
        \midrule
        between sync                & $-0.08^{*}$ & $-18.97^{*}$ & $-16.68^{*}$ \\
        within sync                 & $0.05$      & $13.21$      & $19.11^{*}$  \\
        per-capita income           & $0.06^{*}$  & $7.78$      & $13.78^{**}$ \\
        \midrule
        $\mathbf{R^2_{adj}}$ & $\mathbf{0.54}$ & $\mathbf{0.43}$ & $\mathbf{0.42}$ \\
         \bottomrule
    \end{tabularx}
    \label{table:2-hours}
\caption{Regression results obtained with a 2-hours time window. As for statistical significance of the beta weights, we use the following notation: *$p<0.05$, **$p<0.01$.}
    \end{table}

\begin{table}[ht]
    \centering
    \footnotesize

    \begin{tabularx}{\columnwidth}{@{}Xrrrrr@{}}
        \toprule
        & \textbf{Referendum turnout} & \textbf{Blood donations} & \textbf{Association Density} 
        \\
        \midrule
        between sync    & $-0.09^{*}$  & $-21.15^{*}$ & $-16.44^{*}$   \\
        within sync     & $0.08$       & $17.51$      & $23.26^{**}$   \\
        pop. illiterate & $-0.06^{**}$ & $-8.46$     & $-17.26^{**}$ \\

        \midrule
        $\mathbf{R^2_{adj}}$ & $\mathbf{0.61}$ & $\mathbf{0.50}$ & $\mathbf{0.69}$ \\
         \bottomrule
    \end{tabularx}
     \caption{Referendums turnout, Blood donations, Association density represented by \emph{between} and \emph{within synchronization}, controlled for population illiteracy. As for statistical significance of the beta weights, we use the following notation: *$p<0.05$, **$p<0.01$.}
   \label{table:pop-illiterate}
\end{table}

A set of additional regression analyses tests whether a different covariate, i.e., the fraction of illiterate population, can substitute per-capita income in the regression model.  
Table \ref{table:pop-illiterate} shows the obtained results. It is interesting to see that the use of this covariate does not change the basic structure of our regression: positive correlation with \emph{within synchronization}, negative correlation with \emph{between synchronization}, although statistical significance is weaker than in the case of per-capita income. This can be partially explained by the low number of data-points, which can influence the p-value.

\end{appendices}

\bibliographystyle{unsrt} % or try abbrvnat or unsrtnat
\bibliography{socialk.bib}

\end{document}